\documentclass[10pt]{iopart}
\usepackage{epsfig}
\usepackage{iopams}

\def\lesssim{\mathrel{\hbox{\rlap{\hbox{\lower4pt\hbox{$\sim$}}}\hbox{$<$}}}}
\def\gtrsim{\mathrel{\hbox{\rlap{\hbox{\lower4pt\hbox{$\sim$}}}\hbox{$>$}}}}

\def\lsim{\mathrel{\hbox{\rlap{\hbox{\lower4pt\hbox{$\sim$}}}\hbox{$<$}}}}
\def\gsim{\mathrel{\hbox{\rlap{\hbox{\lower4pt\hbox{$\sim$}}}\hbox{$>$}}}}

\usepackage{ulem}

\newcommand{\s}{\;\mathrm{s}}

\newcommand{\km}{\;\mathrm{km}}
\newcommand{\yr}{\;\mathrm{yr}}

\newcommand{\Msol}{\mathrm{M}_{\odot}}

\newcommand{\Mch}{\mathcal{M}}

\newcommand{\apj}{ApJ}
\newcommand{\apjl}{ApJ}

\newcommand{\aap}{A$\&$A}
\newcommand{\araa}{ARAA}

\newcommand{\mnras}{MNRAS}
\newcommand{\pasj}{PASJ}
\newcommand{\prd}{PRD}
\newcommand{\aj}{AJ}
\newcommand{\nat}{Nature}

\def\beq{\begin{equation} }
\def\eeq{\end{equation} }
\def\spose#1{\hbox to 0pt{#1\hss}}
\def\ltsim{\mathrel{\spose{\lower.5ex\hbox{$\mathchar"218$}}\raise.4ex\hbox{$\mathchar"13C$}}}

\def\spose#1{\hbox to 0pt{#1\hss}}
\def\lta{\mathrel{\spose{\lower 3pt\hbox{$\mathchar"218$}}
        \raise 2.0pt\hbox{$\mathchar"13C$}}}
\def\gta{\mathrel{\spose{\lower 3pt\hbox{$\mathchar"218$}}
        \raise 2.0pt\hbox{$\mathchar"13E$}}}
\def\newblock{\hskip .11em plus .33em minus .07em}

\begin{document}

\title[Electromagnetic signatures of PTA sources]{Electromagnetic signatures of supermassive black hole
binaries resolved by PTAs}

\author{Takamitsu L. Tanaka$^{1}$ and Zolt\'an Haiman$^2$}

\address{$^1$ Max Planck Institute for Astrophysics, Karl-Schwarzschild-Str.~1, D-85741 Garching, Germany\\
$^2$ Department of Astronomy, Columbia University, New York, USA}

\ead{taka@mpa-garching.mpg.de, zoltan@astro.columbia.edu}

\begin{abstract}
  Pulsar timing arrays (PTAs) may eventually be able to detect not
  only the stochastic gravitational-wave (GW) background of SMBH
  binaries, but also individual, particularly massive binaries whose
  signals stick out above the background.  In this contribution, we
  discuss the possibility of identifying and studying such
  ``resolved'' binaries through their electromagnetic emission.  The
  host galaxies of such binaries are themselves expected to be also
  very massive and rare, so that out to redshifts $z\approx 0.2$ a
  unique massive galaxy may be identified as the host. At higher
  redshifts, the PTA error boxes are larger and may contain as many as
  several hundred   massive-galaxy interlopers.  In this case, the
  true counterpart may be identified, if it is accreting gas
  efficiently, as an active galactic nucleus (AGN) with a peculiar
  spectrum and variable emission features.  Specifically, the binary's
  tidal torques expel the gas from the inner part of the accretion
  disk, making it unusually dim in X-ray and UV bands and in broad
  optical emission lines.  The tails of the broad wings of any
  FeK$\alpha$ emission line may also be ``clipped'' and missing.  The
  binary's orbital motion, as well as the gas motions it induces, may
  trigger quasiperiodic variations. These include coherent flux
  variability, such as luminous, multi-wavelength flares, as well as
  Doppler shifts of broad emission lines and ``see-saw'' oscillations
  in the FeK$\alpha$ line.
  Additional features, such as evidence for a recent major merger
  or dual collimated jets, could also corroborate the counterpart.   These properties would make resolved PTA
  sources stand out among AGN with similar overall luminosities and
  allow their identification.
\end{abstract}

\maketitle

%To turn comments on-and-off in DVI/PDF:
%1. Replace all: "" with "%\txmod"
%    Replace all: "" with "%\txmvb"
%    Replace all: "" with "%\txmvb"
%    Replace all: "%" with "%\txmva"

\section{Introduction}
\label{sec:intro}

Multi-messenger observations of SMBH binaries---that is, synergistic
studies of both their electromagnetic (EM) and gravitational-wave (GW)
emission---represent an extraordinary astronomical opportunity to
probe SMBH accretion, general relativity and cosmic expansion (see
reviews by \cite{Haiman+09, Schnittman11,Dotti+2012}).  To date,
theoretical studies of multi-messenger SMBH astronomy have
centered around massive binaries whose coalescences can be observed by
a future space-based laser interferometer such as \textit{eLISA}
\cite{eLISA}.  In this paper, we discuss the prospects of
multi-messenger astronomy for SMBHs that are individually detected
(``resolved'') by pulsar timing arrays (PTAs) because their GW signals
stick out above the stochastic $\sim{\rm nHz}$ background
 \cite{SVV09, SesVec10, CorCor10, DengFinn11, BabakSesana12, Petiteau+13} .
PTA sources differ from \textit{eLISA} sources in that the former are much more
massive (total mass $M\gta 10^{8-9}\;\Msol$ as opposed to $\sim
10^{5-7}\;\Msol$), nearby (redshifts $0.1\ltsim z\ltsim 1$
vs. redshifts up to $z\sim 20$), and are most likely to be ``caught''
well before their coalescence. In fact, the vast majority of PTA
sources will not merge within a human lifetime.

Given that most galaxies appear to have a massive nuclear BHs
\cite{KR95}, the formation of SMBH binaries is an inevitable
consequence of the hierarchical structure formation paradigm, in which
galaxies are built up by mergers between lower-mass progenitors.  The
generic expectation is that merger events deliver the two nuclear
SMBHs \cite{Springel+05,Robertson+06}---along with stars and
gas \cite{BarnesHernquist91}---to the central regions of the new
post-merger galaxy.
 A close BH pair (separated by a $\sim$kpc) is formed,
surrounded by a stellar bulge and dense nuclear gas
\cite{Begelman+80, Yu02}, subsequently decaying its orbit and becoming
a bound binary.  The nuclear gas is expected to cool rapidly, and
settle into a rotationally supported circumbinary disk
\cite{Barnes2002,Escala+05b}.  Such a gas disk can serve the dual
purpose of promoting the binary's orbital decay and producing EM
emission in the form of a luminous active galactic nucleus (AGN).

There are large theoretical uncertainties, regarding both the
formation of compact SMBH binaries and their EM signatures.  The
``final parsec'' problem---whether the binary can coalesce in a Hubble
time \cite{Begelman+80}---is still an open issue for the most massive
($\gsim 10^9~{\rm M_\odot}$) BH pairs.  The efficacy of gas in
bringing the binary to the compact, GW-emitting regime remains unclear
\cite{Escala+05b,Mayer+07, Callegari+09, Colpi+09,
  Hayasaki09,Lodato+09, Nixon+11}, especially in light of the fact
that the massive disk required may be gravitationally unstable for
these BHs \cite{Goodman03, Cuadra+09, HKM09}.  On the other hand,
recent $N$-body simulations suggest that stellar scatterings may
harden SMBH binaries more efficiently than was previously recognized
\cite{Berczik+06,Preto+2011}.  At the sub-parsec separations
relevant to GW emission, open questions include (i) the expected
amount and distribution of gas in the binary's vicinity, (ii) the
coupled dynamical evolution of the binary+gas system, and (iii) the
generation of radiation and radiative transfer effects
(i.e., AGN physics in general ).

Despite these uncertainties, theoretical studies have converged on a
relatively simple picture for the behavior of a SMBH binary embedded
in a thin, prograde, co-planar circumbinary disk.  In this paper, we
apply this picture to assess the prospects for uniquely identifying EM
counterparts of binaries resolved by PTAs.  We focus our attention on
two distinct theoretical expectations that make these prospects
favorable.  The first is the fact that, being the most massive SMBHs
in the local Universe, resolved PTA sources are likely to reside in
host galaxies that are very massive and rare, which will greatly
narrow the search for candidate host galaxies in the detection error
box.\footnote{The flip-side of this is the relatively low expected
  number---a few tens---of resolved PTA binaries.}  The second is
that an AGN accretion disk is predicted to have qualitatively
different geometrical and dynamical properties when the central object
is a prograde binary SMBH instead of a solitary SMBH.  We will detail
the various persistent and time-variable EM signatures that have been
predicted for accretion flows onto pre-coalescence SMBH binaries.

The remainder of this paper is organized as follows.  In
\S~\ref{sec:hosts}, we discuss the likely properties of host galaxies
of PTA sources, and estimate the number counts of hosts within the
expected detection error volume.  In \S~\ref{sec:evol}, we review the
current understanding of the orbital evolution of SMBH binaries, with
an emphasis on the role and distribution of a circumbinary accretion
disk.  An overview of the EM signatures proposed for accreting,
compact SMBH binaries are presented in \S~\ref{sec:EMsigs}.  We
offer our brief conclusions in \S~\ref{sec:concl}.

\section{Demography of Resolved  PTA Sources and Their Host Galaxies}
\label{sec:hosts}

According to population synthesis studies \cite{SVV09, CorCor10,
  KocSes11, Sesana+12}, binaries individually detectable by PTAs will
have large chirp masses $\Mch\equiv
M_{1}^{3/5}M_{2}^{3/5}M^{-1/5}=\eta^{3/5}M\sim10^{8-9}\Msol$, where
$M_{2}\le M_{1}$ are the masses of the two BHs, and $\eta\equiv
M_{1}M_{2}/M^{2}\le 1/4$ is the symmetric mass ratio.  The PTA
binaries are also relatively nearby, with their redshift distribution
declining steeply outside the range $0.1\ltsim z\ltsim 1.5$ (the
cutoffs are due to the small local volume at low $z$, and to the
attenuation of the GW signal at high $z$).  Finally, resolved PTA
binaries are expected to have periods of $P\sim 0.1-10\yr$.  They are
expected to be---almost by definition, because they rise above the
stochastic GW background---the most massive and nearly equal-mass
(i.e. $M_1\sim M_2$) SMBH binaries in the local Universe.
 
It is well known that the masses of nuclear SMBHs correlate with the
velocity dispersion $\sigma$
\cite{FerrareseMerritt00,Gebhardt+00,Tremaine+02,Gultekin+09,
  Graham+11} and the luminosity $L$
\cite{KR95,Magorr+98,MarconiHunt03,Lauer+07,Graham07} of the host
galaxy.  More massive SMBHs also reside in more massive dark matter
halos \cite{Ferrarese02, Dutton+10}, although this correlation may be
less tight  \cite{KorBen11}. Thus, the exceptional masses of
resolved PTA sources imply that their hosts should be either giant
elliptical galaxies or among the most massive spiral galaxies (with
$\sigma\gsim 200 \km\s^{-1}$ of the spheroid component
\cite{Gultekin+09}).  It is also reasonable to expect the host galaxy
to be the product of a relatively recent major merger.  Based on the
fact that cosmological simulations exhibit a weak environmental
dependence of the major merger rates of halos on overdensity
\cite{Fakhma09, Bonoli+10}, \cite{TMH12} concluded that PTA sources
are more likely to reside in a field galaxy than in a cluster.
Finally, it is plausible that a PTA source is more likely to be
shining as a luminous AGN than the average SMBH, because (i) major
galaxy mergers have been associated with luminous quasar activity
 \cite{Sanders+88,Hernquist89,Carlberg90}, and (ii) the presence
of gas may be instrumental in bringing the binary to small separations
where they emit GWs (see \S~\ref{sec:evol} below).  Because the
Eddington ratio of luminous AGN tends to peak around a value $\sim
0.2$ \cite{Kollmeier+06}, one could also look for counterparts
by focusing on the most luminous AGN in the error volume (as opposed
to all galaxies).
 
\begin{figure}[t]
\begin{center}
  \includegraphics[height=0.35\textheight]{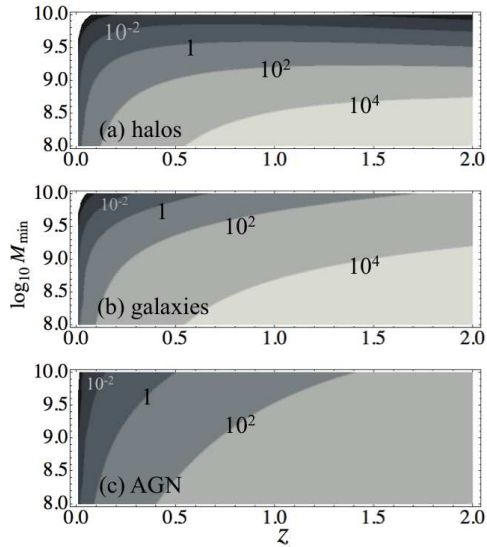}
%  \plotfiddle{fig1.eps}{1.4in}{0.}{54.}{54.}{-140}{-260}
\end{center}
\caption{\label{fig:hosts}
\footnotesize
Estimated number counts in a PTA detection error volume
($\Delta \Omega \sim 3\;\deg^2$, luminosity distance error $20\%$; \cite{CorCor10})
of (a) dark matter halos, (b) galaxies and (c) AGN that are massive/luminous enough
to host a SMBH with minimum mass $M_{\rm min}$.
SMBH-host correlations and mass/luminosity functions
are drawn from previous studies as described in the text.
In this idealized estimate, the number of plausible host candidates (interlopers)
may be sufficiently low for $M\gsim 10^9\;\Msol$ binaries at $z\ltsim 1$
to allow for the unique identification of the EM counterpart.   }
%\vspace{-1\baselineskip}
\end{figure}

The fact that their hosts should (on average) also be massive and rare
objects can be used to the observer's advantage in the search for EM
counterparts.  Figure~\ref{fig:hosts} (adopted from \cite{TMH12})
shows the approximate counts of host galaxy interlopers---i.e., the
number of dark matter halos, galaxies and AGN that are massive or
luminous enough to plausibly host a nuclear SMBH binary of mass
$M_{\rm min}$---in the PTA detection error volume estimated by
\cite{CorCor10}.  The number of massive halos was calculated using the
$M_{\rm BH}-M_{\rm halo}$ relation of \cite{Dutton+10} and the halo
mass function of \cite{Jenkins+01}; the luminous galaxies using the
$M_{\rm BH}-L_{\rm gal}$ relation of \cite{Lauer+07} and the
luminosity function of \cite{Gabasch+06}; and luminous AGN using the
luminosity function of \cite{Hopkins+07b} while assuming a
conservative minimum Eddington ratio of 0.01.  We note that the
estimate of interloping AGN counts is based on optical luminosity
functions. This is important, because as we will discuss below in
\S~\ref{sec:EMsigs}, the PTA sources may have unusually weak UV and
X-ray signatures.

The error volume consists of a luminosity distance uncertainty of $20\%$
and an angular uncertainty of $\Delta \Omega = 3\;{\rm deg}^2$.
This is an optimistic estimate, in that it assumes that the distances
to the pulsars in the array can be used to locate the source
(see also \cite{DengFinn11, BabakSesana12, Petiteau+13}).
Figure \ref{fig:hosts} suggests that in a random volume of such a size,
the number of luminous/massive interloping galaxies
are at most a few dozen if $M\gsim 10^9$, and of order one or less
for a wide range in redshift.
More conservative error boxes  \cite{SesVec10}
result in $\sim 10$ times more interlopers, but the expected number
is still of order unity or less for the most massive ($>10^9\Msol$) SMBHs
at $z\lsim 0.5$ \cite{TMH12}.
These estimates suggest that in the best-case scenario,
the error volumes of resolved PTA sources may contain one
or a handful of massive galaxies or luminous AGN that can harbor them.
At worst, narrowing the search field to the most luminous hosts
could facilitate the identification effort.
Additional considerations, such as morphological evidence
of a recent major merger, may also serve as a ``smoking gun'' feature
that gives away the host galaxy. 

However, Figure~\ref{fig:hosts} also shows that in many cases,
the number of host-galaxy interlopers could be in the hundreds.
Furthermore, SMBH-host correlations are trends only,
and trends have outliers \cite{vandenBosch+12}.
It is also not out of the question that SMBH binaries and their host galaxies
may preferentially lie on extreme tails of known SMBH-host correlations.
Therefore, additional observable signatures of PTA sources are desirable,
both to corroborate the tentative identification of the EM counterpart
when the number of ``plausible'' hosts is one,
and to positively identify the true host when the number of
interloping luminous galaxies is large.

\section{Co-evolution of SMBH Binaries and Their Circumbinary Disks}
\label{sec:evol}

In this section, we briefly review the current theoretical understanding
of how SMBH binaries evolve to the small separations where
they emit GWs and can be detected by PTAs.
This will serve to lay out a basic framework for the gaseous
environment of PTA sources, which will in turn inform
predictions for their EM signatures (\S~\ref{sec:EMsigs}).

\begin{figure}[t]
\begin{center}
  \includegraphics[height=0.35\textheight]{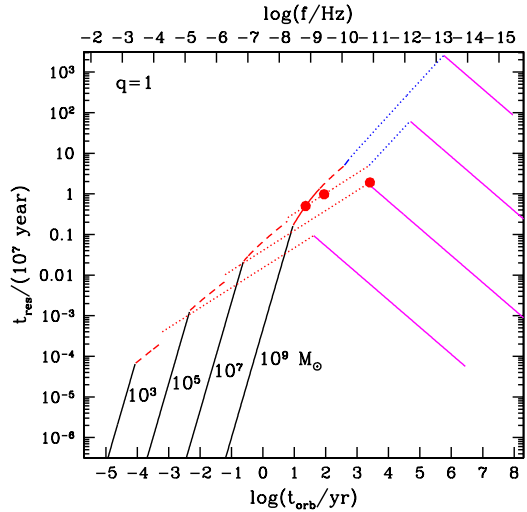}
%  \plotfiddle{fig1.eps}{1.4in}{0.}{54.}{54.}{-140}{-260}
\end{center}
\caption{\label{fig:tres}
\footnotesize
Residence times $t_{\rm res}=-a/(da/dt)$ for
equal-mass SMBH binaries as a function of orbital time $t_{\rm orb}$.
The orbital frequency corresponding to the same separation $a$ is
shown on the top horizontal axis.  The four curves correspond to binaries
with total masses of $M=10^3,~10^5,~10^7$, and $10^9~{\rm M_\odot}$,
as labeled. The large red dots denote the critical radius beyond which
the assumed circumbinary accretion disk is unstable to fragmentation
(Toomre parameter $Q<1$). In each case, the solid black lines at the
highest frequencies denote the regime where the orbital decay is
dominated by GW emission. The blue and red lines correspond to the
regime where the orbital evolution is driven by the torques due to the
circumbinary disk: blue (red) color indicates that the disk mass
enclosed within the binary's orbit is larger (smaller) than that mass
of the secondary.  The dotted/dashed/solid portion of each curve
indicates the three distinct outer/middle/inner radial zones in the
accretion disk. Finally, the magenta lines at the lowest frequencies
indicate the regime where the binary's orbital decay is driven by a
surrounding spherically symmetric stellar background.}
%\vspace{-1\baselineskip}
\end{figure}

During and following a galaxy merger, the two nuclear SMBHs sink to
the center of the post-merger potential well through dynamical
friction, and form a bound binary.  Minor mergers of galaxies are
unlikely to produce SMBH binaries, as the smaller galaxy can be
tidally stripped and dynamical friction will be unable to deliver its
SMBH to the center of the post-merger galaxy
\cite{Callegari+09}.  Therefore, astrophysical SMBH binaries
are expected to consist of BHs of comparable masses ($0.01 \lsim q <
1$ \cite{Sesana+12}.  The binary's subsequent orbital decay is
illustrated in Figure~\ref{fig:tres} (adapted from \cite{HKM09}),
which shows the theoretical \textit{residence time} $t\equiv
-a/(da/dt)$---where $a$ is the binary's semimajor axis---for
equal-mass binaries of various masses, as a function of their orbital
time.  As the binary evolves from large to small separations, it
undergoes three distinct regimes of evolution.
 
{\em Stellar Scattering.}  At large separations, the orbital decay is
driven by three-body interactions with stars that wander into the
binary's orbital path.  The residence time in this regime is shown by
the solid magenta lines in Figure~\ref{fig:tres}.  These curves, which
employ equation~10 from \cite{Cuadra+09} based on the results on
\cite{MM03}, as well as the $M-\sigma$ relation from
\cite{Ferrarese02}, assume that the loss cone is refilled by stars
diffusing on the steady-state, two-body relaxation timescale in a
spherically symmetric potential. These curves are conservative; recent
$N$-body simulations have shown that time-dependent, rotating,
triaxial or axisymmetric stellar potentials can accelerate binary
evolution substantially \cite{Berczik+06, Preto+2011}.

{\em Binary-Disk Angular Momentum Exchange}.  At smaller separations,
the binary can accelerate its orbital evolution by exchanging orbital
angular momentum with a surrounding gas disk.  This interaction also
distorts the surface density profile of the disk near the binary's
orbit.  Over these long timescales, this interaction is expected to
align the binary and the disk, so that they become prograde and coplanar
\cite{IPP99}.  In the mass-ratio regime of interest
($q\gsim0.01$), the secondary is massive enough to open a low-density
annular gap about its orbital path.  The secondary is essentially a
particle in the disk, and migrates inward on the viscous diffusion
timescale (so-called ``Type II'' migration).  The dotted blue lines in
Figure~\ref{fig:tres} show the residence times in this regime.  At
sufficiently small orbital radius, the local disk mass becomes too
small to substantially absorb the orbital angular momentum of the
secondary, slowing its migration.  The red curves in Figure
\ref{fig:tres} show the residence times in this regime.  In standard
models of geometrically thin disks \cite{SS73}, this stalling
occurs at fairly large orbital separations, typically several thousand
Schwarzschild radii.  Finally, the outer disk may be unstable against
gravitational fragmentation \cite{Goodman03,HKM09}; critical
values of $a$ for gravitational stability are marked with red dots in
Figure \ref{fig:tres}.  As the figure shows, this is a particular
problem for the most massive ($\gsim 10^9~{\rm M_\odot}$) binaries,
whose stable disk may extend only over a factor of few in radii.

{\em GW Emission.} Finally, once the binary is sufficiently compact,
its orbit will decay via GW emission. The black curves in Figure
\ref{fig:tres} show the evolution in this regime, using the quadrupole
formula for circular binaries \cite{Peters64}.  The evolution in this
final regime is strongly self-accelerating, with $t_{\rm res}\propto
a^4$.
 
Figure~\ref{fig:tres} also shows the three zones of a standard thin
accretion disk, depending on the dominant source of opacity (free-free
in the outer region; electron-scattering otherwise) and pressure
(radiation in the inner region; gas pressure otherwise).

The distribution and behavior of gas in the immediate vicinity of the
BHs are two crucial factors that determine the EM signatures of the
binary. Unfortunately, the details are the most murky for the compact
systems of interest here. In the late stages of disk-driven
migration, when the secondary is more massive than the local disk
mass, the system is unable to maintain steady-state Type II
migration.  The evolution of the binary and the disk become strongly
coupled.  Because numerical simulations cannot cover the time-spans
necessary to study the system from disk-binary coupling to merger,
much of the current theoretical understanding of binary evolution in
this regime comes from analytic and 1D calculations
\cite{SyerClarke95,IPP99,AN02,Lodato+09,LiuShap10,KHL2012a,KHL2012b,Rafikov2013}.

The gas interior to the secondary's orbit, where the viscous time is
short, is believed to drain onto the primary, creating a central
circumbinary cavity \cite{MP05}.  As gas continues to accrete inward
from larger radii, it is pushed outward by the binary's tidal torques
and piles up outside a radius $R\sim 2a$ \cite{ArtLub94}, much like a
``dam'' in a river \cite{Pringle91,SyerClarke95,IPP99,Tanaka11}.  Such
inner cavities are seen in many numerical simulations, beginning with
the seminal work by \cite{ArtLub94}.  However, in the absence of
comprehensive numerical calculations, it is unknown whether cavities
remain open throughout the binary's evolution.  Ongoing accretion may
refill the cavity if the ``dam'' is leaky---or, even if the ``dam'' is
initially effective, continued accumulation of gas may eventually
cause it to ``burst'' and fail (but this may not occur in the most
massive binaries) \cite{KHL2012a,KHL2012b}.  If the ``dam'' holds,
then its location is expected to follow the binary's orbital
decay---the damming comes at the expense of the binary's orbital
angular momentum---i.e.  always remain at the resonance radius $\sim 2a$,
at least until the binary becomes GW-driven and decouples from the disk
(see below).
 
Simulations also show that even when binary torques open a cavity, the
``dam'' is generally porous.  Modulated by the binary orbit,
circumbinary gas leaks periodically into the cavity, in narrow,
elongated streams
\cite{ArtLub96,Hayasaki+07,MM08,Cuadra+09,Roedig+11,Shi+12}. The rate
at which mass ``leaks'' into the cavity is of order $\sim 10-100\%$ of
the accretion rate in the circumbinary disk far outside the cavity
\cite{Ochi+05,DOrazio+2013}.

Thus, the strongly coupled evolution and behavior of the binary and
the disk remains a highly nontrivial and open theoretical problem,
despite the considerable progress made by both semi-analytic and
numerical (including GR and/or MHD) calculations.  Ironically, for
GW-emitting binaries, an additional complication is posed by the fact
that the binary \textit{decouples} from the disk.  Because GWs cause
the binary's orbital decay to accelerate as $da/dt\propto a^{-3}$, the
binary will eventually harden faster than the surrounding gas can
viscously respond, and begin to outrun it \cite{MP05}.  Even when the
binary and disk become thus decoupled, a significant amount of gas may
be able to follow the binary as it hardens (note that the torques
holding the ``dam'' recede, as the binary accelerates; \cite{TM10, TMH12}).
Simulations in the relativistic regime found that gas
streams can follow the binary to small radii (several gravitational radii,
$R_{\rm g}\equiv GM/c^2$;
 \cite{Noble+12,Farris+11,Farris+12}), although these simulations
start with small binary separations and assume that the gas is present
near the binary a those initial separations.  Nevertheless,
considering that prior to becoming GW-driven the gas density outside
the ``dam'' could have been very high due to pile-up, it is plausible
that even GW-driven binaries are surrounded by copious amounts of
accreting gas.

Finally, since the circumbinary gas can accelerate the evolution of
the binary, it can reduce the total expected number of PTA sources. As
Figure~\ref{fig:tres} shows, at PTA frequencies, gas is more important
for lower-mass binaries. This produces the happy result that the
unresolved GW background (composed of the emission from $\Mch\lsim
10^8~\Msol$ binaries) is reduced, while the more massive,
individually resolvable binaries are largely unaffected and therefore
stand out at higher signal-to-noise ratio \cite{KocSes11}.
  
Because the innermost regions of accretion flows are where most of the
luminosity and broad emission lines of AGN are generated, an empty or
low-density cavity would have significant consequences for the
observational appearance of an accreting SMBH binary.  Similarly, any
time-dependent dynamical features of the accreting gas driven by the
binary's orbital motion is likely to produce conspicuous time-variable
signatures.  In the next section, we turn to the specific EM
signatures that have been proposed.

\section{Electromagnetic Signatures}
\label{sec:EMsigs}

 Here, we review the various EM signatures that have been predicted for
compact SMBH binaries that may be individually resolved by PTAs.  All
of these signatures assume that the binary is accreting gas from large
radii at a rate comparable to a ``normal'' AGN (i.e. close to the
Eddington rate), and rely on the basic geometrical and dynamical
features of the accretion flow that distinguish these binaries from
solitary SMBHs.  The underlying features fall roughly into three
categories: (i) imprint of the binary on the distribution of gas in
its neighborhood or the geometry of its accretion flow; (ii) dynamics
of the accreting gas, as driven by the binary's time-dependent
potential; and (iii) the orbital motion of the binary itself.

\subsection{Signatures due to the Cavity}

The presence of a central, empty (or nearly empty) cavity directly
implies that the binary would be deficient in high-energy thermal
photons.  To see this, one can naively compute the AGN spectrum for a
standard thin disk that is truncated inside the ``dam'' radius
$R\approx 2a$. The spectral energy distribution (SED) of such a disk
peaks at a frequency corresponding to the black-body temperature at 
this truncation radius,
\beq
\nu_{\rm peak}\sim 10^{15}
\left(\frac{M}{10^9\;\Msol}\right)^{1/4}
\dot{m}^{1/4}
\left(\frac{P}{1\; {\rm yr}}\right)^{-1/2} \; {\rm Hz},
\eeq
which is much smaller than $\nu_{\rm peak}\sim 10^{16}-10^{17}~{\rm
  Hz}$ in a comparable accretion disk around a solitary SMBH of the
same mass \cite{TMH12,GultekinMiller12,Tanaka2013}.  Above, $\dot{m}$
is the accretion rate outside the cavity in units of the critical rate
corresponding to the Eddington luminosity.  This estimate remains
valid even when accounting for the surface density pile-up, until the
very final stages before the merger, when GW emission has caused the
binary to completely leave the circumbinary disk behind \cite{TMH12}.
Gas inside the cavity, as well as the additional heating caused by the
pile-up, can compensate somewhat for the high-energy decrement in the
SED, but this extra heating is significant only for lower ($\Mch\lsim
10^8 {\rm M_\odot}$) binaries \cite{KHL2012b}, and not enough to shift
the peak frequency \cite{Lodato+09}.

\begin{figure}[t]
\begin{center}
  \includegraphics[height=0.35\textheight]{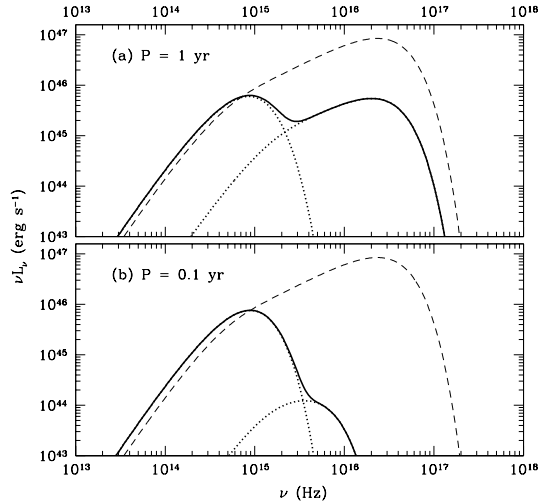}
%  \plotfiddle{fig1.eps}{1.4in}{0.}{54.}{54.}{-140}{-260}
\end{center}
\caption{\label{fig:SED}
\footnotesize
Estimated SEDs of circumbinary disks around a SMBH binary with total
mass $M=10^9\;\Msol$ and mass ratio $M_1:M_2=4:1$.
The dotted curves show the emission from the circumbinary disk
truncated by a central cavity
(bump at $\nu\sim 10^{15}~{\rm Hz}$)
and from the circumsecondary disk (higher-frequency bump).
The solid curves show the composite spectrum,
and the dashed curve shows, for comparison, the
SED of an Eddington accretion disk around a single SMBH
of the same total mass.
}
%\vspace{-1\baselineskip}
\end{figure}

Figure \ref{fig:SED} (adopted from \cite{TMH12}; see also \cite{MP05,
  TM10, Tanaka2013}) shows the SEDs from (i) the circumbinary disc
(dotted lines, low-frequency bump) alongside (ii) a circumsecondary
disc (high-frequency bump) with a size equal to the Hill stability
radius.  The binary in this example has a total mass $10^9 \;\Msol$,
mass ratio $q=1/4$ and orbital periods of $1~{\rm yr}$ (top panel) and
$0.1~{\rm yr}$ (bottom).  The circumbinary gas profile around the
GW-driven binary was calculated from the point of decoupling, using a
1D diffusion equation with a moving inner boundary condition for the
``dam.''  The surface density of the circumsecondary disk is
calculated by assuming that it is fueled via leakage of gas into the
cavity, at a rate that is $10\%$ of the circumbinary gas supply rate.
The circumbinary gas density is enhanced by a factor of $3$ to show
the effect of gas pile-up.  The composite SEDs are shown by solid
lines, and the SED from a disk (with no pile-up)
around a single SMBH of the same total
mass is shown by the dashed curves for comparison.

\begin{figure}
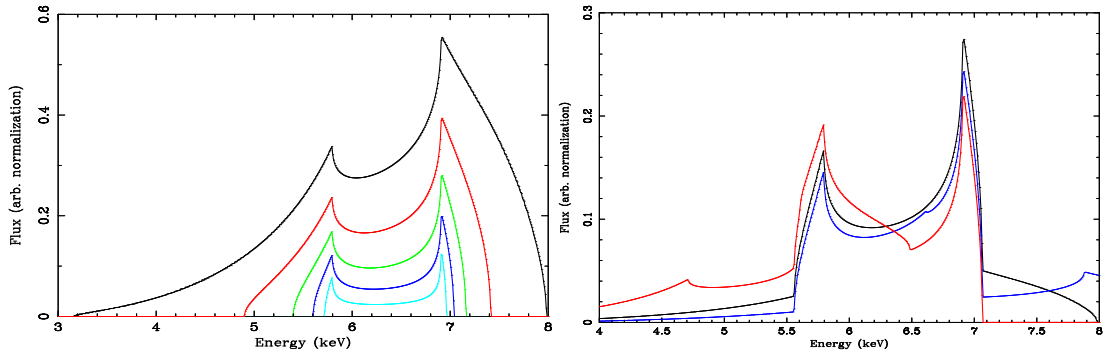

\includegraphics[width=0.35\textwidth,height=0.35\textheight,angle=-90]{fig4a.eps}
\includegraphics[width=0.35\textwidth,height=0.35\textheight,angle=-90]{fig4b.eps}
\caption{\footnotesize
The change in the broad FeK$\alpha$ line profile due to a central cavity in a
  circumbinary accretion disk.  
{\em Left panel: ``clipped wings'' due to central cavity.}
The broad FeK$\alpha$ emission is
  assumed to originate from the region of a standard accretion disk inside $100~R_{\rm g}$, with an
  $r^{-2.5}$ X-ray emissivity profile, and the disk is viewed at an
  inclination angle of $\theta=60^{o}$. The solid black curve shows
  the ``usual'' line profile for a single BH, with the inner disk extending
  from $100~R_{\rm g}$ inward to $6~R_{\rm g}$.  Other curves assume the disk
  has a central cavity, and the gas distribution extends inward only to $20~R_{\rm g}$
  (red curve), $40~R_{\rm g}$ (green curve), $60~R_{\rm g}$ (dark blue curve),
  or $80~R_{\rm g}$ (light blue curve).  As the size of the central cavity increases,
  the broad component of the line decreases in magnitude and its blue
  and red wings are increasingly suppressed.
{\em Right panel: ``see-saw wings'' due to circumsecondary's orbital
  motion.} The black curve shows the Fe~K$\alpha$ emission line from
the main circumbinary disk ($55-100~R_{\rm g}$; i.e. in-between the green
and dark blue curves in the left panel) plus a weak secondary broad
component ($10\%$ of the intensity of the full disk profile) due to an
accretion disk around the secondary black hole, located at $30~R_{\rm g}$,
centered on the line centroid energy (6.40keV).  The red curve shows
the effect of shifting the centroid of the weak secondary component
red-ward to $5.2$keV. The blue curve shows the effect of shifting the
centroid of the weak secondary component blue-ward to $7.3$keV. The
progression from red curve to blue curve occurs over half the orbital
time of the binary. Observationally, we expect a 'see-saw' oscillation
between the blue and red wings of the line over the binary's orbit.
The curves in both panels are binned at approximately the energy
resolution ($\sim 7$eV) expected for \textit{Astro-H}
\label{fig:FeKa}}
\end{figure}

Apart from the conspicuous depression of the flux in the UV and X-ray
bands, Figure \ref{fig:SED} shows that for moderately wide ($P$ of
several years) and/or distant ($z\gsim 0.5$) binaries, the downturn in
the spectrum could be observed even at optical wavelengths.  These
features may cause SMBH binaries to be misidentified as quiescent
galaxies, because they may be missed in color- and X-ray-selected
samples.  The presence of the cavity would also suppress UV and
optical broad emission lines \cite{TMH12}, as well as the X-ray
FeK$\alpha$ line.  In Figure~\ref{fig:FeKa}, we show an illustration
of the latter: the left panel shows the expected line profile (adapted
from \cite{FeKa2013}) for a solitary BH with the accretion disk
extending from 6-100 gravitational radii (black curve) and for binary
BHs that carve out a central cavity of various sizes (other, colored
curves).
 For a fiducial disk with viscosity parameter $\alpha=0.01$
and scale-height $(H/r)=0.1$ around a  $10^8\;\Msol$ SMBH ,
we expect the profile to progress from the solid black curve
to the solid blue curve on a timescale of $\sim 150$ yrs. These ``clipped wing'' features will therefore be stable
(unevolving) for massive PTA binaries, and possibly observable for
binaries caught at small enough separations ($\lsim 100$ gravitational radii)
where the FeK$\alpha$ line is detectable.

As mentioned above, the cavity may not be completely empty, because
the gas inside the annular gap may not have been completely depleted
\cite{Lodato+09, Chang+10}, or because gas leaking into the cavity may
fuel small ``mini-disks'' around one or both SMBHs
\cite{Hayasaki+08,Sesana+12,Farris+12}.
In general, gas inside the cavity can
produce high-frequency photons, but not enough to shift the peak
in the thermal SED; the decrement is still significant \cite{TMH12,GultekinMiller12}.
In very late stages of merger, gas inside the cavity may be tidally heated
by the shrinking binary, resulting in enhanced high-frequency
emission \cite{AN02,Chang+10} (although unfortunately, this effect is
expected to be weak; see \cite{Baruteau+12}).  For systems that are on
the verge of merger or that have recently merged, we may witness the
birth of a bright quasar: the SED of the outer disk may harden
gradually and monotonically, as the cavity fills in, with brightening
optical and UV emission, over several decades \cite{THM10}.

\subsection{Dynamics: Variability due to Periodic Accretion into the Cavity}

The gas that leaks into the cavity may trigger periodic luminosity
variations on either the orbital timescale \cite{ArtLub96, HKM09}
   or its harmonics \cite{Sesana+12,DOrazio+2013}.  The
periodic mass supply onto the circumprimary and circumsecondary
mini-disks can cause periodic fluctuations in luminosity
\cite{Hayasaki+08, Cuadra+09}.  The streams may also shock the
mini-disks \cite{TMH12}, or be flung back out and shock the cavity
wall \cite{DOrazio+2013}.  These fluctuations may be detectable with
high-cadence, long-term monitoring, but a major challenge will be to
disentangle them from intrinsic AGN variability \cite{Ulrich+97,
  Gaskell+06}.  Here a possibility is to look for the presence of
harmonics: in particular, for an unequal-mass binary
(with, say $q\approx 0.3$), a periodogram
of the mass accretion rate into the central regions of the cavity
shows strong spikes at both the binary's orbital time $t_{\rm orb}$,
as well as at $0.5t_{\rm orb}$ (and also at the orbital period at the
cavity wall, $(3-4)t_{\rm orb}$) \cite{DOrazio+2013}.  Seeing periodic
variability on two timescales with a 1:2 ratio (plus a longer
timescale) could therefore be a ``smoking gun'' for the presence of a
binary.

Recently, \cite{Tanaka2013} proposed that periodic streams into the
cavity may trigger much more prominent variability than previously
thought.  The tidal elongation of the streams would cause the leading
end to collide with the trailing portions shortly after pericenter,
and shock-heat.  Because the streams are highly eccentric and have
large kinetic energies relative to the circular-orbit value at
pericenter ($\ltsim 0.1a$ from simulations of
\cite{Hayasaki+08,Cuadra+09}), the shock can circularize the gas
orbits about the nearby SMBH and heat it to nearly virial
temperatures.  Such a hot, optically thick accretion flow could
accrete on timescales shorter than a binary orbit (provided that the
pericenter of the stream is less than the Hill radius of the nearby
SMBH), and fuel transient, luminous flares instead of long-lasting
mini-disks.  Interestingly, such binary-induced flares---rapidly
decaying, multi-wavelength flares with AGN-like SEDs peaking in the
UV---may closely resemble tidal disruptions of stars.  However,
because PTA sources are too massive to cause tidal disruption events,
a measurement of the SMBH mass in the flaring system should
distinguish between the two flare mechanisms.  The observation of
repeating flares in the same galaxy (on the timescale of a few years,
matching the GW frequency) could also corroborate a SMBH binary
origin, since the tidal disruption rate is thought to be only $\sim
10^{-5}~{\rm yr}^{-1}$ in typical galaxies.
(The observation of multiple (genuine)
stellar tidal disruptions in a short span
may also indicate the presence of a SMBH binary
\cite{Chen+09,Chen+11}). 

Quasi-periodic luminosity fluctuations have been observed in several
AGN, most notably in OJ 287, a BL Lac object that exhibits pairs of
optical flares---brightening by as much as 5 magnitudes---every
$\approx 11.7$ years, going back to the late 19th century.  Other,
less prominent examples include 3C345 \cite{LobanovRoland05}
and Mrk 501 \cite{Roedig+09}.  OJ 287 has long been speculated
to be a SMBH binary \cite{Sillanpaa+88}.  In recent years,
\cite{Valtonen+06, Valtonen+08, Valtonen+12} have modeled the system
as an extremely massive eccentric SMBH binary ($M_1=1.8\times
10^{10}\;\Msol$ and $M_2=1.4\times 10^8\;\Msol$) with a tilted
circumprimary accretion disk (without a cavity).  In this
interpretation, the flares are caused by an eccentric secondary
impacting the disk twice per orbit. Alternatively, \cite{Tanaka2013} suggests
that the presence of a cavity and the self-shocking of the
periodically leaked gas may explain both the outbursts and the
intervening relative quietude without invoking an ultra-massive
primary.  Because the disk-impacting model makes highly precise
predictions of the timing of future flares (based on the relativistic
precession of the binary), whereas the cavity-flare model of
\cite{Tanaka2013} predicts the flares are only quasi-periodic (with
scatter expected in orbital dynamics of the gas leaking into the
cavity), future observations could distinguish the two mechanisms.  If
OJ 287 indeed contains a binary, it would be the closest known object
to a source of detectable low-frequency GWs.  If the masses in the
disk-impacting model of \cite{Valtonen+12} are correct, then OJ 287
is unlikely to be detectable with PTAs, as the mass ratio is too small
to produce a resolvable GW signal.  However, strong modulations of the
accretion streams were found \cite{DOrazio+2013} to require a more
equal-mass binary ($q\gsim 0.05$) which could be detectable by PTAs.

\begin{figure}[t]
\begin{center}
  \includegraphics[height=0.35\textheight]{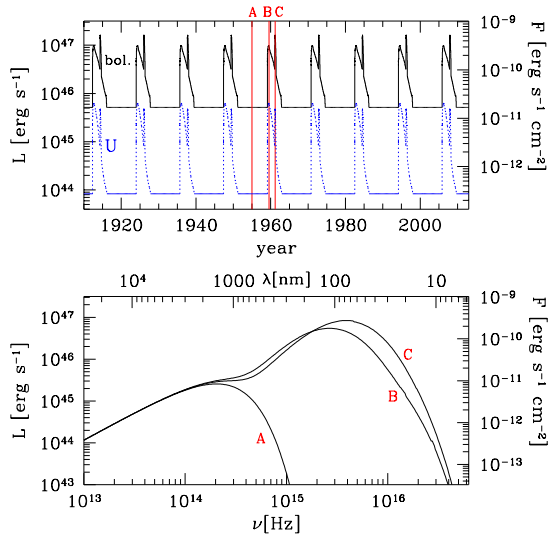}
%  \plotfiddle{fig1.eps}{1.4in}{0.}{54.}{54.}{-140}{-260}
\end{center}
\caption{\label{fig:flare} \footnotesize Toy light curves and SED
  snapshots for a flaring SMBH binary with $M=10^9 \;\Msol$,
  $M_2/M_1=1/4$, $z=0.3$ and $P_{\rm obs}=11.7~{\rm yr}$.  The
  parameters were chosen to roughly correspond to the known properties
  of OJ 287.  The model assumes one-dimensional accretion of an
  advective torus that has been shock-heated to its virial
  temperature.  }
%\vspace{-1\baselineskip}
\end{figure}

Figure \ref{fig:flare} shows a toy flare model for OJ 287 by
\cite{Tanaka2013}.  The binary is modeled as having a total mass
$M=10^9 \;\Msol$ and a $1:4$ mass ratio, lying at $z=0.3$ with an
observed period of $11.7~{\rm yr}$.  A fraction of $5\%$ of the mass
supply rate of the outer disk is assumed to accrete onto the SMBHs.
The top panel shows the variation over time of the bolometric
luminosity and the U-band flux.  The bottom panel shows snapshots of
the SED (at points marked in the light curves of the top panel), in
the observer's rest-frame.  The model is highly idealized, assuming
viscosity prescription similar to the $\alpha$ model \cite{SS73} and
treating the flaring accretion flow as a one-dimensional advective
torus.  Both the viscous spreading and the dynamics of such accretion
flows are unknown, and future numerical simulations will be required
to more robustly predict its emission signatures.

\subsection{Kinematics: Features that Vary with the Binary Orbit}

 If gas accretes onto both SMBHs, this may generate
a pair of collimated jets \cite{Palenz+10b, Moesta+12}
that prescess and vary with the binary orbit \cite{OShaughnessy+11}.
Dual jets could also leave behind peculiar radio-lobe morphologies
as fossil imprints of the preceding histories of accretion,
orbital motion and evolution \cite{Roos+93, Liu+03, GopalKrishna+03}
and spin evolution \cite{MerrittEkers02, Bogdanovic+07}
histories of the binary.
(However, radio lobe morphologies have alternative explanations
\cite{Kraft+05, SaripalliSub09}).
 
Finally, as mentioned above, the gas leaking into the cavity may fuel
small ``mini-disks'' around one or both SMBHs
\cite{Hayasaki+08,Sesana+12,Farris+12}.  If these disks are stable and
long-lasting (i.e. persist over many binary periods), then
time-variable signatures could be produced by their
kinematics---i.e. the changing Doppler shifts along the black holes' orbits.

For example, if both BHs have individual FeK$\alpha$ lines (produced
by fluorescence of their individual hot X-ray coronae), then there
would be a pair of emission lines \cite{Sesana+12}, shifting in
frequency in tandem, and in opposite directions.  In principle, it
will be feasible, with a next generation X-ray observatory with high
spectral resolution (such as Athena) to identify such shifting double
FeK$\alpha$ line features.    Even if only the secondary BH has its
own mini-disk and FeK$\alpha$ line, it could produce measurable
``See-saw oscillations' in the blue and red wings of the line \cite{FeKa2013}.
To illustrate this latter possibility, in the right panel Figure~\ref{fig:FeKa},
we show the spectrum of a binary system, which has a ``naked'' primary
(without a disk), but a secondary with a mini-disk (located at $30~R_{\rm g}$),
as well as the circumbinary disk.  As the figure shows, as the secondary
is moving towards the observer vs. receding, its emission adds to the
blue vs. red wing of the overall line.

In principle, periodic oscillations---or at least shifts---could be
visible for broad optical and UV lines, as well, provided that at
least one of the BHs has its own broad line region
\cite{ShenLoeb2010,Montuori+2012,Dotti+2012}.
A search for such frequency shifts has been performed recently among 88 quasars,
 selected from the SDSS, based on offsets between their broad lines and
the quasar's rest-frame. Follow-up spectroscopy has revealed several
candidates with frequency shifts, but further observations are needed
to test the binary hypothesis in each case.  Although it has turned
out difficult, in general, to prove the presence of a binary using
only EM observations, any observed frequency shift will be useful as
corroborating evidence once combined with the GW
observations---especially if the tentative EM periodicity for exactly one of the
sources in the GW error box matches the GW period.

\section{Conclusions}
\label{sec:concl}

 Discovering the EM emission from a massive PTA binary BH would be
revolutionary in several ways, providing unique probes of SMBH
accretion astrophysics, general relativity and cosmic expansion.  At a
more basic level, EM and GW observations could reinforce each other,
and together would likely provide a much better evidence for the
presence of a compact SMBH binary than either observation by itself.
There are several plausible EM signatures, based on the presence of a
cavity in the inner $\sim 100 ~R_{\rm g}$ around the binary, and on
time-variability in the continuum and line emission, linked to the
binary period. We have highlighted these in this paper, while also
emphasizing the large uncertainties that currently exist in the
theoretical expectations.

The most robust signatures of a PTA binary is likely to be based on
variability.  With periods of months to years, the search for periodic
EM signals from massive binaries in the PTA range is well suited to
forthcoming forthcoming surveys, such as the \textit{Large Synoptic
  Survey Telescope} in the optical and \textit{eROSITA} in the X-rays.
Given the projected time-line of improvements to PTAs (leading up to
the the Square Kilometer Array) that will allow them to resolve
individual SMBH binaries, it is plausible that convincing candidate
PTA binaries could be first identified in such EM surveys
\cite{HKM09}.  In this case, the pre-cataloged massive binaries could
aid in the detection of GWs from these objects, rather than the other
way around.

\section*{Acknowledgments}

We thank Alberto Sesana and Constanze R\"odig for useful discussions,
and our collaborators Kristen Menou, Barry McKernan, and Bence Kocsis
for permission to draw on joint work.  ZH acknowledges  support from
NASA grant NNX11AE05G.

\bibliographystyle{unsrtnat}

\end{document}